\documentclass{elsarticle}

 \usepackage{graphicx}
\usepackage{amssymb}

\journal{Powder Technology}

\begin{document}

\begin{frontmatter}

 \title{Morphology and displacement of dunes in a closed-conduit flow \tnoteref{label_note_copyright} \tnoteref{label_note_doi}}

\tnotetext[label_note_copyright]{\copyright 2016. This manuscript version is made available under the CC-BY-NC-ND 4.0 license http://creativecommons.org/licenses/by-nc-nd/4.0/}

\tnotetext[label_note_doi]{Accepted Manuscript for Powder Technology, v. 190, p. 247-251, 2009, http://dx.doi.org/10.1016/j.powtec.2008.04.065}

\author{E.M. Franklin\corref{cor}}
\ead{franklin@fem.unicamp.br}
\author{F. Charru}
\ead{charru@imft.fr}
\cortext[cor]{Corresponding author.}
\address{Institut de M\'ecanique des Fluides de Toulouse, All\'ee du Pr. Camille Soula, 31400, Toulouse, France}



%\thanks[label2]{}


%\title{Morphology and displacement of dunes in a closed-conduit flow}


%\author{}

%\address{}

\begin{abstract}
% Text of abstract
The transport of solid particles entrained by a fluid flow is frequently found in industrial applications. A better knowledge of it, is of importance to improve particle related industrial processes. When shear stresses exerted by the fluid on the bed of particles are bounded to some limits, a mobile layer of particles known as bed-load takes place in which the particles stay in contact with the fixed bed. If it takes place over a non-erodible ground, and if the particle flow rate is small enough, an initial thin continuous layer of particles becomes discontinuous and composed of isolated dunes.
	  We present here an experimental study to understand some features of the dynamics of isolated dunes under a fluid flow using a closed-conduit experimental loop made of transparent material. Acquired data concerns mainly dune morphology and displacement velocity under different conditions: different types of beads (diameters and densities) and different water flow conditions.
	  We observed that an initial pile of beads placed in the conduit is rapidly deformed by the water flow, adopting a ``croissant'' shape, like barchan dunes found in deserts at a much larger scale.  We observed also self-similarity in dunes dimensions and that dune displacement velocity scales with the inverse of their dimensions. The variation of the dune displacement velocity with the fluid shear velocity is discussed here.
\end{abstract}

\end{frontmatter}

% main text
\section{Introduction}
\label{intro}

The transport of solid particles entrained by a closed-conduit fluid flow is frequently found in industrial applications. It is found, for example, in the petroleum, food and pharmaceutical industries. Concerning the size of the solid particles, several orders of magnitude may be present: from some tens of microns to some tens of millimeters. A better knowledge of this kind of transport and of its associated pressure drop is then of great importance to improve particle related industrial processes. Nevertheless, up to now, it has not been theoretically well understood.

When shear stresses exerted by the fluid flow on the bed of particles are able to move some of them, but are relatively small compared to particles apparent weight, a mobile layer of particles known as bed-load takes place in which the particles stay in contact with the fixed bed. The thickness of this mobile layer is a few particle diameters. Bed-load existence depends on the balance of two forces:
a) an entraining force of hydrodynamic nature, proportional to $\tau d^{2}$, where $\tau$ is the bed shear stress and $d$ is the mean particle diameter;
b) a resisting force related to particles apparent weight, proportional to $(\rho_{s} - \rho)gd^{3}$, where $\rho$ is the fluid density, $\rho_{s}$ is the particles density and $g$ is the gravitational acceleration.

One important dimensionless parameter is then the Shields number $\theta$, which is the ratio of the entraining force to the resisting force:

\begin{equation}
        \theta = \frac{\tau}{(\rho_{s}-\rho)gd}
\end{equation}

Bed-load takes place for $0.01 \, \lesssim \, \theta \, \lesssim \, 1$.

Under the fluid flow, the plane bed may become unstable and deformed, generating dunes. So, if bed-load takes place over a non-erodible ground, as a closed-conduit wall for instance, and if the particle flow rate is small enough, an initial thin continuous layer of particles becomes discontinuous and composed of isolated dunes. In a closed-conduit those isolated dunes generate supplementary pressure loss. Moreover, as they migrate inside the closed-conduit, they may generate pressure fluctuations.

We present here an experimental study concerning the latter situation. The objective is to understand some features of the dynamics of isolated dunes, notably their displacement and deformation. To the authors knowledge, this situation, although very common in industrial applications, has not yet been studied. The next section describes the experimental set-up. It is followed by three sections describing the experimental results: the dune morphology (section \ref{morph}), the shear velocity estimation (section \ref{vitfrot}) and the dune velocity (section \ref{vit}). Follows the conclusions section (section \ref{conc}).

\section{Experimental set-up}
\label{set-up}

A closed-conduit experimental loop of rectangular cross-section (for simplicity, we name it ``channel'' in the following) and made of transparent material was used to investigate the dynamics of isolated dunes under a water flow.

Concerning the fluid flow, we are interested here in the fully turbulent regime. This can be defined in terms of the Reynolds number based on the cross section average velocity $\bar{U}$ and on the conduit height $H$: 

\begin{equation}
  Re=\frac{\bar{U}H}{\nu} \, > \, 10000
\end{equation}

Where $\nu$ is the kinematic viscosity. In the turbulent case, the shear velocity $u_{*}$ is defined by $\tau \, = \, \rho u_{*}^{2}$.

In our experiments we used water as the fluid media and some spherical beads as granular media: glass beads with density $\rho_{s} \, = \, 2500 kg/m^{3}$ and mean diameter $d \, = \, 0.50 \, mm$, $d \, = \, 0.20 \, mm$ and $d \, = \, 0.12 \, mm$ and zirconium beads with density $\rho_{s} \, = \, 3800 kg/m^{3}$ and mean diameter $d \, = \, 0.19 \, mm$. The water flow rate was varied between $6 \, m^{3}/h$ and $10 \, m^{3}/h$, which gives the following range of Shields number $\theta$ and Reynolds number $Re$: $0.02 \, < \, \theta \, < \, 0.25$ and $13000 \, < \, Re \, < \, 24000$.

In order to have a good control of dune displacement and deformation under a permanent water flow, it is desirable to have low turbulence levels at the channel inlet. This was achieved by establishing a gravitational flow by means of a constant level head tank rather than the direct use of a pump.

\begin{figure}
    \centering
        \includegraphics[width=\columnwidth]{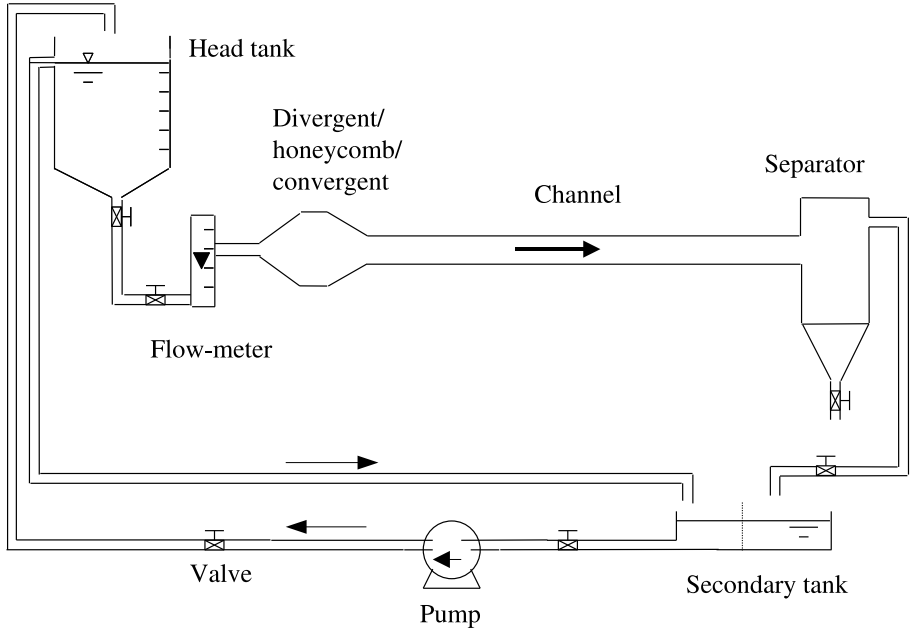}
    \caption{Experimental equipment.}
    \label{fig:schema}
\end{figure}

The experimental loop is made of (figure \ref{fig:schema}): 

\noindent 1) A head tank. This constant water level tank gives a $2 \, m$ head pressure at the channel (test section). Water is continuously pumped to the head tank (from the secondary tank) and the level is assured constant by an overflow passage (discharging in the secondary tank).\\
\noindent 2) An electromagnetic flowmeter, which measures the fluid flow rate in the channel.\\
\noindent 3) A divergent/honeycomb/convergent device, which can break large turbulent structures.\\
\noindent 4) A channel (test section).\\
\noindent 5) A fluid-particles separator. Particles settle due to a strong expansion of the fluid flow.\\
\noindent 6) A secondary tank. The channel fluid flow and the head tank overflow are discharged in this tank.\\
\noindent 7) A water lifting pump, which continuously pumps water from the secondary tank to the head tank. The pumped water flow rate is larger than the water flow rate in the channel, maintaining the head tank water at a constant level.\\
\noindent 8) Valves to control the water flow rate.\\

The channel is a six metre long horizontal closed-conduit of rectangular cross-section ($120 \, mm$ wide by $60 \, mm$ high), made of transparent material. One of the advantages of the rectangular cross-section channel in comparison with a tube is its plane horizontal bottom surface. The fluid flow in this kind of channel is well-known \cite{Melling,Leutheusser}.

A mirror inclined at $45^{o}$ made it possible to use a single camera to obtain top and side views of isolated dunes. The camera, mounted on a rail system, was above the channel and had a direct top view of the dune. The mirror, close to one of the vertical sides, indirectly provided the dune side view to the camera. As the camera was mobile and a rule was fixed on the channel, we could follow each isolated dune and record its displacement and deformation.

The experimental procedure was as follows: a conical pile of beads was built, from a funnel, in the channel (already filled with water). The funnel was located at $4.15 \, m$ from the channel inlet. Then, a constant fluid flow was established. The conical pile rapidly became a dune and its motion was recorded by a camera. With this procedure, each experiment concerns one single isolated dune.

\section{Dune Morphology}
\label{morph}
As soon as the fluid flow attains the desired flow rate, solid particles in the surface of the pile move as bed-load, while particles inside the pile stay at rest. Before being displaced over a measurable distance, the pile is deformed and adopts a ``croissant'' shape as shown in figure 2, like barchan dunes found in deserts at a much larger scale \cite{Bagnold,Elbelrhiti,Kroy}, but also on Mars in a still larger scale \cite{Claudin}. In our experiments, barchan dunes had a length of a few tens of millimetres while desert barchans have a length of a few tens of metres.

We observed that the barchan dune is displaced by an erosion-deposition mechanism: erosion upstream and deposition downstream. Nevertheless, this mechanism is quite complex. Upstream the dune crest, in the region called ``upwind side'', there is solid particles erosion and migration to the crest. Once near the crest, the particles settle due to the flow perturbations caused by the dune shape (including a water recirculation zone downstream the crest). There is then particle accumulation near the crest, which gives avalanches on the face downstream to the crest, known as ``lee side''. Due to the water recirculation zone downstream from the crest, the particles in the front of the dune are not entrained downstream, except at the horns, where the recirculation zone is weaker. So, downstream from the crest, erosion exists only at the horns. Differently from the erosion over the upwind face, particles eroded over the horns are entrained far from the dune. This kind of granular dynamics was observed for desert barchan dunes \cite{Kroy} and aquatic barchan dunes in a quite different configuration from our experiments \cite{Hersen}.

In our experiments, there is no solid particles flux delivered upstream of the barchan dune. Thus, the net mass balance is negative due to particle erosion at the horns, this means that the dune size decreased during migration. Nevertheless, we observed that the dune moves whilst always maintaining the barchan shape.

\begin{figure}[h]
    \centering
        \includegraphics[width=\columnwidth]{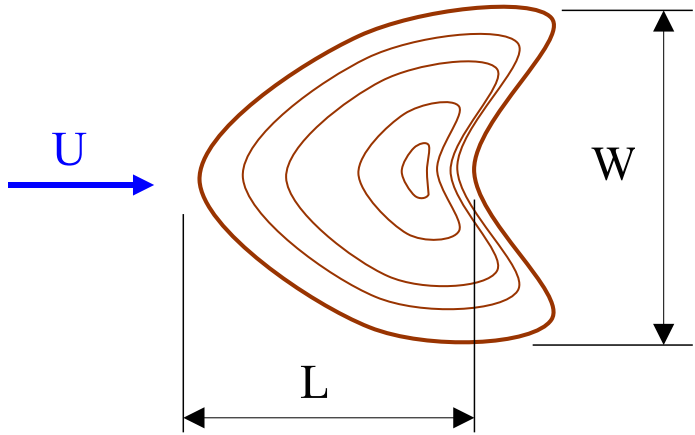}
    \caption{Top view of a barchan dune (scheme).}
    \label{fig:dune}
\end{figure}

A diagram of the barchan dune top view is shown in figure \ref{fig:dune}, which defines its dimensions.  $L$ is the distance, in the symmetry plane, from the front (downstream) to the rear (upstream) of the dune, not taking into consideration the horns. $W$ is the dune width. $h$, not shown in this figure, is the dune height.

\begin{figure}[h]
    \centering
        \includegraphics[width=\columnwidth]{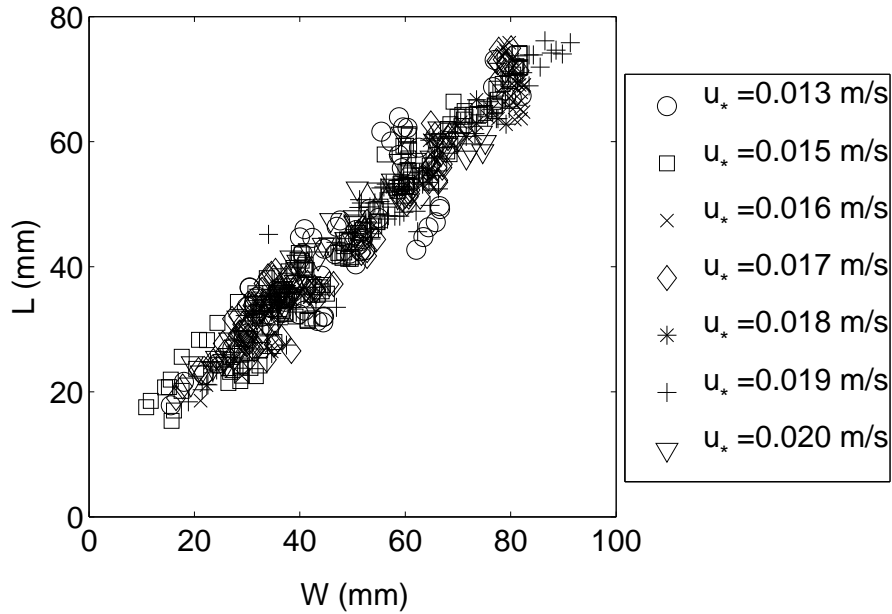}
    \caption{Dunes length $L$ to width $W$ ratio. Markers do not distinguish between the four types of beads (glass beads with $d \, = \, 0.50 \, mm$, $d \, = \, 0.20 \, mm$ and $d \, = \, 0.12 \, mm$ and zirconium beads with $d \, = \, 0.19 \, mm$ ) because no difference was noted between them.}
    \label{fig:morph}
\end{figure}

Figure \ref{fig:morph} presents the length $L$ to width $W$ ratio of $532$ measurements. The shear velocities indicated in the legend were computed as explained in the section \ref{vitfrot}. This figure shows a remarkably constant $L/W$ ratio: $L \, = \, 0.8W \, + \, 5.7$ (in $mm$). Concerning the other dimensions (figures not shown here), the same kind of linear behavior is found (with only different coefficients), indicating self-similarity in dune dimensions, i.e., roughly constant height, width and length ratios. It can be noted that this has been seen for aeolian dunes and for aquatic dunes in a different configuration \cite{Hersen}. Markers do not distinguish between the four types of beads because no difference was noted.

Those similarities are, in some way, an astonishing result because the bed-load transport mechanism is different for dunes under gas flow from dunes under liquid flow. In air, solid particles move mainly by saltation \cite{Bagnold}, i.e., by ballistic flights, where some particles are ejected by the shock of the falling ones. The air has not so much influence on the particle trajectory other than accelerating it during its flight. When the accelerated particle falls, it can eject other ones. In water, particles move mainly by rolling over the bed or by relatively small jumps. They are dislodged directly by the fluid flow and their trajectory is strongly affected by it.

Thus, the shape similarities between aeolian and aquatic barchans indicate that this shape is independent of the transport mechanism of the solid particles.

\section{Shear velocity $u_{*}$ estimation}
\label{vitfrot}

Melling and Whitelaw \cite{Melling} and Leutheusser \cite{Leutheusser} measured the velocity profiles in a closed-conduit flow of rectangular cross-section in the case of a single-phase flow. They showed that the vertical profile on the vertical symmetry plane of the channel is composed of two logarithmic half-profiles, beginning near the channel horizontal faces and matching each other at the channel axis. In the other vertical planes parallel to the symmetry, they found the same kind of profile, with only slight changes, except very near the vertical lateral walls of the channel (within $5 \%$ of the channel width), where wall and corner effects are stronger.

Between those boundary layers near the vertical lateral walls, the spanwise variation of the longitudinal vertical velocity profile seems very slight. For instance, Melling and Whitelaw \cite{Melling} couldn't determine the variation of the longitudinal shear velocity in the spanwise direction because the variations were of the same order of magnitude as the uncertainty in the measurement (by Laser Doppler Anemometry). In order to obtain an estimation of this variation, they computed the shear stress on the channel walls, $\tau \, = \, \rho u_{*}^{2}$, where they used the mean longitudinal shear velocity (following the channel width), and compared it with the shear stress computed from the channel pressure drop. They found that this latter is $9 \%$ larger than the first one.

In our experiments, the dunes occupied the central part of the channel. Considering the largest dunes, the distance between their borders and the vertical walls was at least of $6 \%$ of the channel width. In general the dunes occupied a narrower region around the vertical symmetry plane. We can consider this region limited to $70 \%$ of the channel width around the vertical symmetry plane. This corresponds to the region occupied by the dunes in $99 \%$ of the experiments.

Based on this, an estimation of the shear velocity can be made by considering a two-dimensional flow and a vertical velocity profile composed of two logarithmic half-profiles. By mass conservation, the integration of the vertical profile equals the flow rate divided by the channel width. Thus, we can estimate the shear velocities $u_{*}$ based on the flow rate given by the flowmeter:

\begin{equation}
        q \, = \,2 \int\limits_{y_{0}}^{H/2} \frac{u_{*}}{ \kappa} \ln \frac{y}{y_{0}} \, dy
	\label{eq:frot}
\end{equation}

where $H$ is the height of the channel, $\kappa \, = \, 0.4$ is the Karman constant, $y$ is the vertical direction and $q$ is the flow rate divided by channel width. $y_{0} \, = \, \frac{\nu}{u_{*}}e^{-5.5 \kappa}$ is a roughness length, defined as the height where the fluid velocity is equal to zero. We estimated the shear velocity from equation (\ref{eq:frot}) with constant $y_{0} \, = \, 10^{-5} \, (m)$, corresponding then to a smooth wall.

\section{Dune Velocity}
\label{vit}
The dune velocity $V$ is computed from the camera images. It is the mean velocity given by the displacement of the front of the dune between two images divided by the time interval between them. This time interval was taken as corresponding to a dune displacement of about one or two times its size. This time interval is sufficient to guarantee a negligible variation of the dune size between the images.

\begin{figure}[h]
    \centering
        \includegraphics[width=\columnwidth]{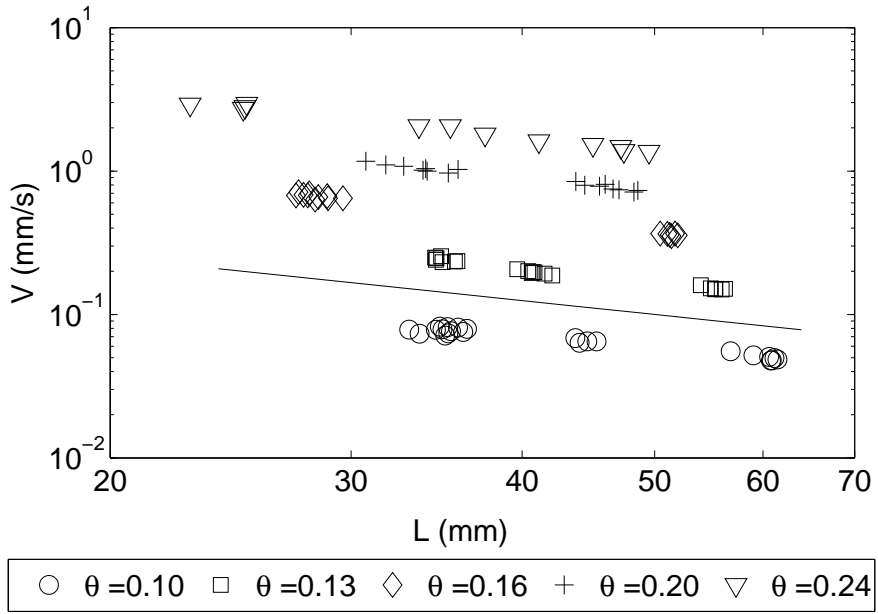}
    \caption{Dune displacement velocity $V$ as a function of their length $L$ for glass beads dunes with $d \, = \, 0.12 \, mm$. Straight-line corresponds to $V \, \propto \, L^{-1}$.}
    \label{fig:vit1}
\end{figure}

Figure \ref{fig:vit1} presents the dune displacement velocity $V$ as a function of their length $L$ for dunes made of glass beads ($\rho_{s} \, = \, 2500 kg/m^{3}$) with $d \, = \, 0.12 \, mm$. The shear velocity $u_{*}$ on the horizontal walls varies between $0.013 \, m/s$ and $0.020 \, m/s$ (fluid flow rate between $6 \, m^{3}/h$ and $10 \, m^{3}/h$) which corresponds to $\theta$ between $0.10$ and $0.24$. Both axis are in logarithmic scale.

\begin{figure}[h]
    \centering
        \includegraphics[width=\columnwidth]{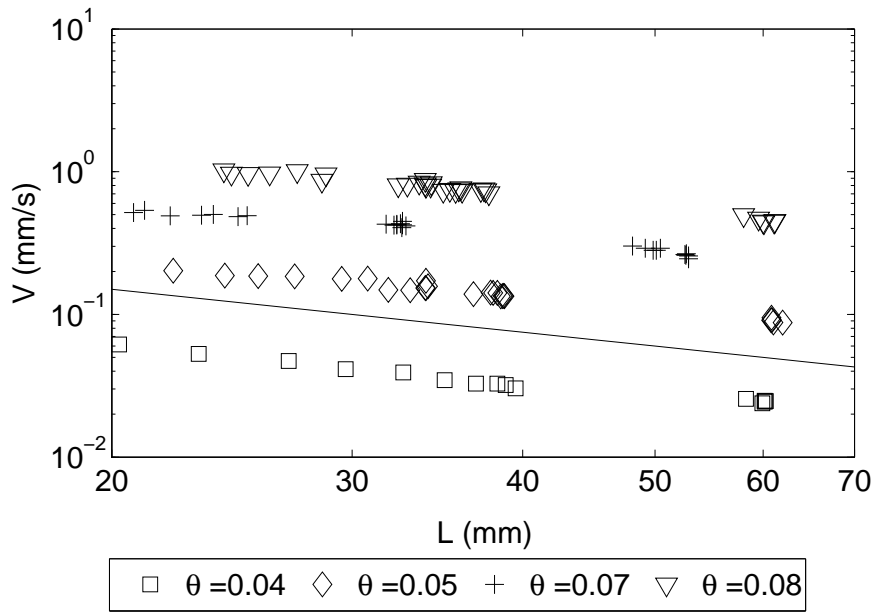}
    \caption{Dune displacement velocity $V$ as a function of their length $L$ for zirconium beads dunes with $d \, = \, 0.19 \, mm$. Straight-line corresponds to $V \, \propto \, L^{-1}$.}
    \label{fig:vit2}
\end{figure}

Figure \ref{fig:vit2} presents the same displacement velocity $V$ for dunes made of zirconium beads ($\rho_{s} \, = \, 3800 kg/m^{3}$) with $d \, = \, 0.19 \, mm$.  The shear velocity $u_{*}$ at horizontal walls varies between $0.015 \, m/s$ and $0.020 \, m/s$ (fluid flow rate between $7 \, m^{3}/h$ and $10 \, m^{3}/h$) which corresponds to $\theta$ between $0.04$ and $0.08$.

For each marker, corresponding to the same friction velocity (or Shields number), we observe some sets of data, each set corresponding to the same dune as it propagates downstream and decreases in size. Considering each type of marker individually, we observe that all the points (and also the sets) are well aligned. Alignment gives us a good indication about the repeatitivity of the tests. The general data alignment indicates that $V$ is a power function of $L$. Also, all lines seem to have the same inclination, $L^{-1}$, present as a reference in the figures, which gives us $V \, \propto \, L^{-1}$. As dune shapes are self-similar, dune displacement velocity is found to be inversely proportional to their size. This functional relation was predicted by Bagnold \cite{Bagnold} and has been observed for aeolian dunes \cite{Bagnold,Elbelrhiti}, but it was not previously measured for barchan dunes in water.

\begin{figure}[h]
    \centering
        \includegraphics[width=\columnwidth]{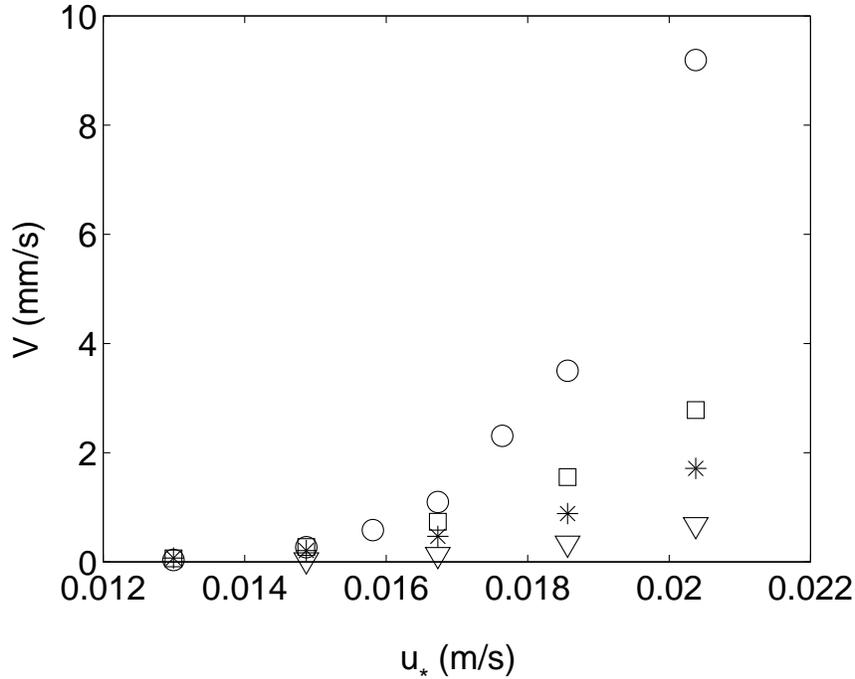}
    \caption{Dune displacement velocity $V$ as a function of the shear velocity $u_{*}$, in the case of a dune with a fixed length $L \, = \, 40 \, mm$. Circles, squares and asterisks correspond to glass beads dunes with $d \, = \, 0.50 \, mm$, $d \, = \, 0.20 \, mm$ and $d \, = \, 0.12 \, mm$, respectively. Triangles correspond to zirconium beads dunes with $d \, = \, 0.19 \, mm$.}
    \label{fig:vit4}
\end{figure}

In our experiments we were able to control fluid flow conditions and we estimated the shear velocities on the channel horizontal walls. To the authors knowledge, this was not done in previous aquatic barchan dune experiments (nor, for obvious reasons, in field measurements of aeolian barchan dunes). For this reason we can present here the variation of the dune displacement velocity $V$ with the shear velocity $u_{*}$. This is shown in figure \ref{fig:vit4} for a fixed dune size ($L \, = \, 40 \, mm$). To draw this graph, it was necessary to interpolate our data because there is no fixed value of $L$ corresponding to the data for all the $\theta$ and bead type range. We interpolated this data following their alignment as $V \, \propto \, L^{-1}$ and we took the values of $V$ corresponding to $L \, = \, 40 \, mm$. Figure \ref{fig:vit4} shows a displacement velocity $V$ variation with the shear velocity $u_{*}$ that seems to follow a power law. Indeed, if we plot it in a logarithmic scale, data seems to be aligned, but inclinations varies between $7$ and $12$. Those exponents are much higher than that involved in the scaling $V \, \propto \, (u_{*})^{3}$ proposed by Bagnold \cite{Bagnold} and usually accepted for the aeolian dunes. Nevertheless care needs to be taken as explained below.

In fact, the values of $u_{*}$ are here close to the threshold $u_{*th}$, otherwise the solid particles would be entrained as a suspension by the liquid flow. So, $u_{*th}$ should be taken into consideration in our case. It is also to be noted that we estimated the shear velocity $u_{*}$ on the channel horizontal walls, while to obtain a scaling as proposed by Bagnold \cite{Bagnold} we would need to use the shear velocity on the dune surface (the Bagnold \cite{Bagnold} scaling uses the bed-load flow rate $Q$ over the dune crest).

The question of the scaling of $V$ with $u_{*}$ for the aquatic barchan dunes in a closed-conduit is still an open problem. Some experiments to determine $u_{*}$ and $u_{*th}$ over the dune surface remain to be performed. Nevertheless we can already state that the threshold shear velocity cannot be neglected in this case, differently from some aeolian cases.

\section{Conclusions}
\label{conc}

The transport of solid particles as bed-load may, when the flux particle discharge is low, give rise to isolated dunes, which are displaced and deformed by the fluid flow. In a closed-conduit, such as found in industrial applications, those isolated dunes generate supplementary pressure loss. Moreover, as they migrate inside the conduit, they may generate pressure fluctuations. A better understanding of dune migration is a key point to control sediment transport.

We have investigated experimentally the displacement and deformation of isolated dunes by a water flow in an horizontal closed-conduit of rectangular cross-section (channel). In each experiment, we placed one single dune in the channel. Dunes were made of beads of several sizes and different materials (glass beads with $d \, = \, 0.50 \, mm$, $d \, = \, 0.20 \, mm$, $d \, = \, 0.12 \, mm$ and zirconium beads with $d \, = \, 0.19 \, mm$) and were subjected to different water flow shear rates ($0.013 \, m/s \, \leqslant \, u_{*} \, \leqslant \, 0.020 \, m/s$, corresponding to $0.02 \, < \, \theta \, < \, 0.25$).

An initial pile of beads placed in the channel is rapidly deformed by the water flow, adopting a ``croissant'' shape, like barchan dunes found in deserts at a much larger scale \cite{Bagnold}, but also on Mars in a still larger scale \cite{Elbelrhiti,Kroy,Claudin}. As the bed-load transport mechanism is quite different in air from in water, these similarities between aeolian and aquatic barchans indicate that their shape is independent of the transport mechanism.

The data concerns dune morphology and displacement velocity. As long as morphology is concerned, we observed self-similarity in dune dimensions, i.e., roughly constant height, width and length ratios, as verified for aeolian dunes and for aquatic dunes in a quite different configuration \cite{Hersen}. Concerning dune displacement velocity $V$, we observed that it is inversely proportional to their size. The scaling of $V$ with the shear velocity $u_{*}$ was not clearly determined and additional experiments remain to be done in order to determine $u_{*}$ and $u_{*th}$ over the dune surface. Nevertheless we show that $u_{*th}$  cannot be neglected in the closed-conduit aquatic case (differently from the aeolian case where $u_{*th}$ may be neglected in the case of very strong winds).

\section{Acknowledgments}
\label{}
Erick de Moraes Franklin is grateful to Brazilian government foundation CAPES for their scholarship support.

% \appendix

% \section{}
% \label{}

%\begin{thebibliography}{00}


% notes:

% \bibitem{label2}
% If there is a note, it should come last:
% \bibitem{label3} \note
% \end{subbibitems}

%\bibitem{Elb} H. Elbelrhiti, S. Douady, B. Andreotti, Field evidence of surface-wave-induced  instability of sand dunes, Nature.  437 (1999) 720-723.

%\bibitem{Bag} R.A. Bagnold, The physics of blown sand and desert dunes, Chapman and Hall, London, 1941.

\nocite{*}
\bibliography{ID15}

\begin{thebibliography}{1}
\expandafter\ifx\csname url\endcsname\relax
  \def\url#1{\texttt{#1}}\fi
\expandafter\ifx\csname urlprefix\endcsname\relax\def\urlprefix{URL }\fi

\bibitem{Melling}
A.~Melling, J.~Whitelaw, Turbulent flow in a rectangular duct, J. Fluid Mech.
  78 (1976) 289--315.

\bibitem{Leutheusser}
H.~Leutheusser, Turbulent flow in rectangular ducts, J. Hydr. Div. 89~(3)
  (1963) 1--19.

\bibitem{Bagnold}
R.~Bagnold, The physics of blown sand and desert dunes, Chapman and Hall,
  London, 1941.

\bibitem{Elbelrhiti}
H.~Elbelrhiti, P.~Claudin, B.~Andreotti, Field evidence for
  surface-wave-induced instability of sand dunes, Nature 437~(04058).

\bibitem{Kroy}
K.~Kroy, S.~Fischer, B.~Obermeyer, The shape of barchan dunes, J. Phys.
  Condens. Matter 17 (2005) S1229--S1235.

\bibitem{Claudin}
P.~Claudin, B.~Andreotti, A scaling law for aeolian dunes on mars, venus, earth
  and for subaqueous ripples, Earth Pla. Sci. Lett. 252 (2006) 30--44.

\bibitem{Hersen}
P.~Hersen, S.~Douady, B.~Andreotti, Relevant length scale of barchan dunes,
  Phys. Rev. Lett. 89~(264301).

\end{thebibliography}
\bibliographystyle{elsart-num}

%\bibitem{}

%\end{thebibliography}

\end{document}